\documentclass[a4paper,fleqn]{cas-dc}

\usepackage[numbers]{natbib}

\usepackage{amsthm}
\usepackage{caption}
\usepackage[position=b]{subcaption}
\usepackage{textcomp}
\usepackage{enumitem}
\usepackage{float}
\usepackage{url}
\usepackage{algorithm}
\usepackage{algorithmic}
\usepackage{tikz,pgfplots}
\usepackage{tikz-qtree}
\usetikzlibrary{trees}
\usepackage[export]{adjustbox}

\pgfplotsset{compat=newest}
\definecolor{mygr}{HTML}{e6e6e6}
\tikzstyle{selected}=[draw,minimum width=1in,text width=2.5in, align=center, anchor=west]

\theoremstyle{definition}
\newtheorem{definition}{Definition}[section]

\begin{document}
\let\WriteBookmarks\relax
\def\floatpagepagefraction{1}
\def\textpagefraction{.001}

% Short title
\shorttitle{Attack Patterns in Cross-chain Bridges}

% Short author
\shortauthors{Samiha, I., Rahman, Y., and Mukta, R.}

% Main title of the paper
\title[mode = title]{Attack Patterns in Cross-chain Bridges: A Trust-based Empirical Analysis}

% First author
\author[1]{Ishraq Samiha}
\fnmark[1]
\ead{ishraqsamiha2000@gmail.com}

% Second author
\author[1]{Yasir Rahman}
\fnmark[1]
\ead{chyyasir2000@gmail.com}

% Third author
\author[1,2]{Rahma Mukta}
\cormark[1]
\ead{rahmamukta@cuet.ac.bd}

% Address/affiliation
\affiliation[1]{organization={Chittagong University of Engineering and Technology},
            city={Chattogram},
            postcode={4349},
            country={Bangladesh}}

\affiliation[2]{organization={Australian National University},
            city={Canberra},
            postcode={2600},
            state={ACT},
            country={Australia}}

% Corresponding author text
\cortext[1]{Corresponding author.}

% Footnote text
\fntext[1]{These authors contributed equally to this work.}

% Here goes the abstract
\begin{abstract}
The rapid growth of multiple blockchain ecosystems has made cross-chain bridges essential for transferring assets and information across independent blockchains. Their multi-component architectures, cross-chain trust assumptions, and concentration of valuable assets expose them to diverse security threats. This paper presents an empirical study of cross-chain bridge attacks based on 48 evidence-supported real-world incidents from January 2022 to August 2026. We propose a behaviour-based attack pattern architecture comprising six trust-violation attack patterns, each defined by the primary trust assumption violated by the attacker: \textit{Proof Forgery}, \textit{Logic Exploitation}, \textit{Event Injection}, \textit{Economic-State Manipulation}, \textit{Credential Seizure}, and \textit{Infrastructure Hijacking}. The architecture further maps these patterns to five functional domains of cross-chain bridges: \textit{Deposit}, \textit{Relay}, \textit{Verification}, \textit{Execution}, and \textit{Governance and Operations}, separating how a trust assumption is violated from where the compromised trust boundary lies. The 48 incidents correspond to \$2.676 billion in gross bridge-specific losses, with the ten largest incidents accounting for 92.4\% of total losses. Measured on the primary loss basis, \textit{Proof Forgery} and \textit{Credential Seizure} make up 86.2\% of all losses, and the \textit{Verification} and \textit{Governance and Operations} domains hold 93.9\%. A handful of very large incidents drives this concentration, yet it still holds when we switch to other loss bases or reclassify the boundary cases. By tying what attackers do to the bridge functions and security assumptions they target, the architecture gives threat modelling for cross-chain bridges an empirical footing.
\end{abstract}

% Keywords


\maketitle

% Main text
\section{Introduction}
\label{sec:introduction}

Blockchain technology now underpins a wide range of applications, including Decentralised Finance (DeFi), digital voting, healthcare, and supply chain management~\cite{crosschain_15,crosschain_16,crosschain_17}. As these applications have different requirements, many independent blockchain networks have emerged, each with its own architecture, rules, and consensus mechanism~\cite{crosschain_21}. This proliferation has created a fundamental interoperability problem: assets and messages recorded on one blockchain cannot be directly assumed to be valid by another. Cross-chain bridges address this problem by enabling assets and information to move between independent networks, but they introduce additional trust assumptions and security-critical components\cite{crosschain_2}.

\begin{figure*}[pos=!tp]
    \centering
    \includegraphics[width=\textwidth,max width=\linewidth]{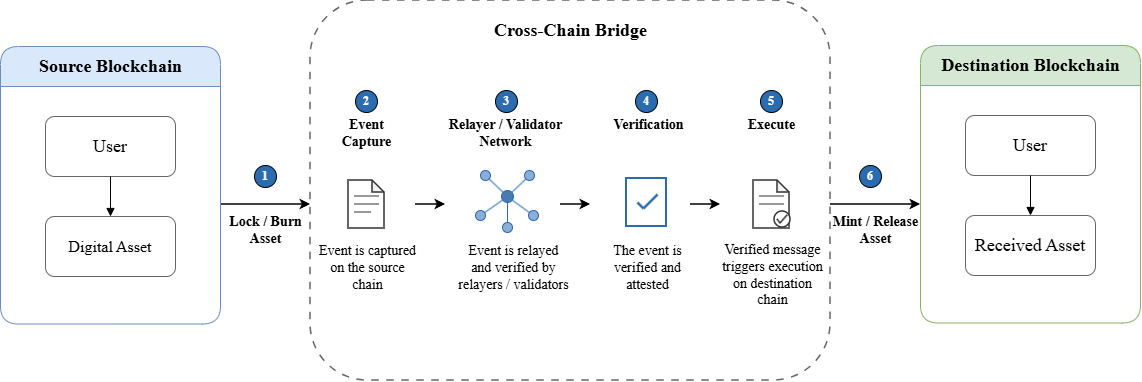}
    \caption{Conceptual overview of a cross-chain bridge illustrating the asset transfer workflow between source and destination blockchains.}
    \label{fig:Conceptual overview}
\end{figure*}

A cross-chain bridge is a protocol that enables the transfer of digital assets and messages between independent blockchain networks, from a source blockchain, where the asset originates, to a destination blockchain, where it is transferred. Although bridges are important for interoperability, they introduce additional attack surfaces and trust assumptions~\cite{crosschain_1}: they often hold large amounts of locked assets and rely on complex architectures with several vulnerable components. Fig.~\ref{fig:Conceptual overview} illustrates a simplified asset-transfer workflow used as the conceptual abstraction in this study. Depending on the bridge design, a user initiates a transfer by locking, burning, or committing an asset on the source side. Relayers or validators then transmit evidence of this event to the destination chain, where it is verified before the destination-side transfer is executed and an equivalent asset is minted or released to the recipient.

Cross-chain bridge attacks have caused significant financial losses, estimated at \$1.5-2 billion in 2022 alone~\cite{crosschain_11}; our dataset's 2022 total of \$1.89 billion (Section~\ref{sec:analysis}) falls within this range. Although bridge exploits accounted for a much smaller share of DeFi losses by 2025~\cite{immunefi2026}, attacks against bridges have continued in 2026. In February 2026, a compromise of the private key controlling the Ethereum-side validator contract of IoTeX's ioTube bridge enabled an attacker to upgrade the contract and drain approximately \$4.3 million in bridge-held assets~\cite{iotex}. In April 2026, a vulnerability in Hyperbridge's Merkle Mountain Range proof-verification logic allowed an invalid proof to be accepted, enabling the unauthorised minting of approximately one billion bridged DOT tokens and resulting in an estimated \$2.5 million in realised losses~\cite{hyperbridge}. Within our dataset, built using the inclusion and loss-measurement criteria defined in Section~\ref{sec:data_collection}, 2026 already accounts for 19 incidents and \$427 million in gross bridge-specific losses by the end of August.

Despite significant advances in cross-chain bridge security, blockchain interoperability remains challenging as attacks continue to evolve in complexity and scale~\cite{crosschain_2}. Existing studies have examined different aspects of bridge security, including vulnerability detection, architectural weaknesses, attack analysis, and runtime monitoring. For example, \textit{XScope}~\cite{crosschain_18} proposed rule-based detection using transaction graph analysis, while subsequent studies investigated software-level vulnerabilities~\cite{crosschain_19}, historical bridge exploits~\cite{crosschain_20}, penetration techniques~\cite{zhao2023comprehensive}, and architectural weaknesses~\cite{crosschain_11}. More recent approaches, including \textit{BridgeGuard}~\cite{crosschain_4}, have explored runtime attack detection. However, these studies generally examine individual vulnerability classes, bridge architectures, detection mechanisms, or selected incidents. Existing studies generally do not provide a unified empirical framework that simultaneously characterises how a bridge trust assumption is violated, locates the compromised functional domain, and measures how these patterns and their financial impact are distributed across real-world bridge incidents.

This paper presents an empirical study that develops and evaluates a behaviour-based attack pattern architecture for cross-chain bridges, aiming to move beyond traditional vulnerability-based classifications toward a better understanding of how attackers exploit the components and operations of bridge systems. The contribution is not that every underlying exploit mechanism is previously unknown. Rather, the study operationalises the violated trust assumption as a common incident-level classification principle, separates it from the compromised functional domain, and evaluates both dimensions empirically across a consistently normalised incident corpus. The analysis follows one conceptual chain: (i) every bridge relies on trust assumptions; (ii) each attack violates one of them through one of six trust-violation attack patterns; (iii) each violation is located in one of five functional domains; and (iv) these patterns and domains are measured across 48 evidence-supported incidents that occurred between January 2022 and August 2026. From this analysis we build a behaviour-based attack pattern architecture in which the six trust-violation patterns fall into three groups and are linked to the five functional domains of a cross-chain bridge. We then measure how patterns, domains, and losses are distributed across the dataset, and characterise the security problem, architectural context, and defensive mitigations of each attack pattern. The key contributions of this paper are summarised as follows:

\begin{itemize}

\item \textbf{Updated Dataset:} We compile 48 evidence-supported cross-chain bridge incidents from January 2022 to August 2026. For every attack it records the gross bridge-specific loss, the root cause, the evidence level, and why the incident was classified as it was. We have released the dataset, codebook, and analysis scripts at \url{https://doi.org/10.5281/zenodo.23271026}.

\item \textbf{Behaviour-Based Attack Pattern Architecture:} We define six trust-violation attack patterns, arranged in three groups: \textit{Proof Forgery (PF)}, \textit{Logic Exploitation (LE)}, \textit{Event Injection (EI)}, \textit{Economic-State Manipulation (ESM)}, \textit{Credential Seizure (CS)}, and \textit{Infrastructure Hijacking (IH)}. Every pattern names the trust assumption the attacker breaks, and explicit rules decide which single pattern an incident belongs to. The patterns overlap with attack classes already known in the literature; what differs is that we group them by the trust they violate, keep them apart from the affected functional domain, and code the whole incident dataset with the same rules.

\item \textbf{Functional-Domain Mapping:} Each attack pattern is placed in one of five cross-chain functional domains, \textit{Deposit}, \textit{Relay}, \textit{Verification}, \textit{Execution}, or \textit{Governance and Operations}, so that it is clear in which part of the bridge the broken trust boundary sits.

\item \textbf{Empirical Analysis:} We examine how patterns, domains, and losses spread across the 48 incidents and change over the years, and we test how sensitive the results are to ambiguous classifications and to the choice of loss basis.

\item \textbf{Attack Pattern Analysis:} For each pattern we describe the mechanism, the domains it affects, how often it occurs, the factors behind it, its implications, the signs that reveal it, and the mitigations that apply. Different bridge attacks can then be compared in the same terms.

\end{itemize}

The rest of the paper is organised as follows. Section~\ref{sec:preli} presents the operational model, the attack-pattern groups, and the functional domains. Section~\ref{sec:related_work} reviews the existing literature. Section~\ref{sec:data_collection} describes how the dataset was built and how incidents were classified. Section~\ref{sec:analysis} presents the empirical analysis of the collected incidents. Section~\ref{sec:proposed_architecture_detailed} presents the proposed behaviour-based attack pattern architecture and describes each attack pattern in detail. Finally, Section~\ref{sec:conclusion} concludes the paper and outlines directions for future work.

\section{Preliminaries}
\label{sec:preli}

This section introduces the operational model, the attack-pattern taxonomy, and the functional-domain model, which together form the proposed behaviour-based attack pattern architecture used throughout the paper.

\subsection{How Cross-Chain Bridges Transfer Assets}

A cross-chain bridge connects two independent blockchains: a source chain $\mathcal{S}$ and a destination chain $\mathcal{D}$. Each chain follows its own consensus rules and maintains its own state, so a blockchain cannot natively assume the validity of state transitions occurring on another chain~\cite{crosschain_18}. Bridge designs vary and include liquidity-based transfers, intent-based designs, and general message passing. For the asset-transfer bridges considered in this paper, two common transfer mechanisms are:

\begin{itemize}
    \item \textbf{Lock-and-Mint (L-M):} The user locks a native asset in a bridge contract on $\mathcal{S}$. Once the lock is confirmed, the bridge creates (mints) an equivalent wrapped token on $\mathcal{D}$~\cite{crosschain_19}.

    \item \textbf{Burn-and-Release (B-R):} To move the asset back, the user destroys (burns) the wrapped token on $\mathcal{D}$. Once the burn is confirmed, the bridge releases the original locked asset on $\mathcal{S}$~\cite{zhao2023comprehensive}.
\end{itemize}

For asset transfers, the bridge should mint or release assets on one chain only when a matching, genuine action has occurred on the other chain~\cite{crosschain_18,crosschain_19}. For general cross-chain messaging, the corresponding security principle is that a destination-chain action should be executed only after a message representing a valid source-chain event has been authenticated.

Both principles rest on a set of trust assumptions. Here, a \emph{trust assumption} is a condition the bridge relies on to authorise a transfer, such as the authenticity of a source-chain event, the validity of a proof, the integrity of an economic state, or control of a privileged credential. Bridge security therefore depends on the correctness of proof verification, the authenticity of cross-chain messages and events, the integrity of the economic state used to approve or price transfers, and the security of privileged credentials and supporting infrastructure~\cite{crosschain_19,crosschain_11}. Every attack in our study breaks one of these assumptions: it either gets the bridge to accept an action that never took place, or it hands the attacker direct control of the locked assets.

\subsection{Groups of Bridge Attack Patterns}

\begin{figure*}[pos=!tp]
    \centering
    \includegraphics[width=\textwidth,max width=\linewidth]{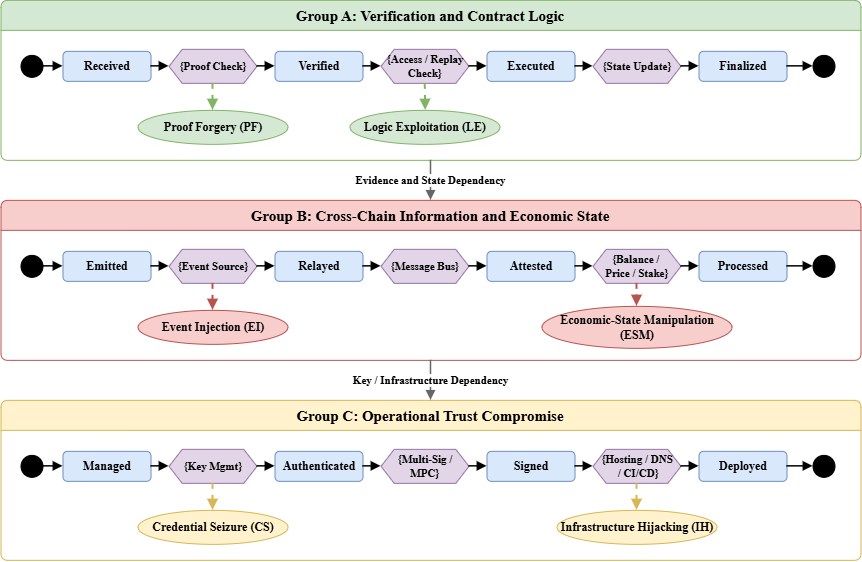}
    \caption{The three groups of attack patterns: Verification and Contract Logic, Cross-Chain Information and Economic State, and Operational Trust Compromise.}
    \label{fig:threemaincatagory}
\end{figure*}

The incidents in our dataset point to six trust-violation attack patterns, each standing for a primary security failure that ends in an unauthorised bridge outcome. We sort them into three categories according to the kind of trust being broken (Fig.~\ref{fig:threemaincatagory}):

\begin{itemize}
\item \textbf{Group A - Verification and Contract Logic:} Trust placed in how the bridge verifies evidence and runs its contract logic.
\begin{itemize}
\item \textit{Proof Forgery (PF):} Forged or invalid evidence gets through because the bridge's verification is defeated or skipped.
\item \textit{Logic Exploitation (LE):} The attacker abuses a contract-logic flaw that lies outside verification, for instance a missing check, a replay, or a faulty upgrade.
\end{itemize}

\item \textbf{Group B - Cross-Chain Information and Economic State:} Trust placed in the information and economic values the bridge relies on.
\begin{itemize}
    \item \textit{Event Injection (EI):} The bridge handles a signal correctly, but the signal itself falsely reports a deposit, burn, or other legitimate event.
    \item \textit{Economic-State Manipulation (ESM):} The attacker distorts an economic value, such as a balance, price, or stake weight, that the bridge depends on when it approves or sizes a transfer.
\end{itemize}

\item \textbf{Group C - Operational Trust Compromise:} Trust in the people and infrastructure operating the bridge.
\begin{itemize}
    \item \textit{Credential Seizure (CS):} The attacker obtains genuine signing, administrator, or validator credentials and uses them for unauthorised actions.
    \item \textit{Infrastructure Hijacking (IH):} The attacker compromises supporting infrastructure, such as hosting, DNS, networking, or build systems, to redirect users or expose credentials.
\end{itemize}

\end{itemize}

Some incidents involve multiple failures. The rules for assigning a single primary pattern are described in Section~\ref{sec:data_collection}.

These groups and patterns describe \emph{what type} of trust assumption is violated and \emph{how} it is violated. To show \emph{where} in the bridge the compromised trust boundary lies, we next divide bridge operation into five functional domains.

\subsection{Five Functional Domains of a Cross-Chain Bridge}

\begin{figure}[pos=!htbp]
    \centering
    \includegraphics[width=0.5\textwidth,max width=\linewidth]{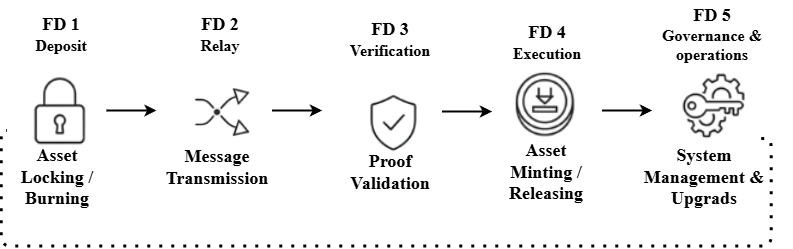}
    \caption{The five FDs of a cross-chain bridge.}
    \label{fig:Functional domain}
\end{figure}

A cross-chain transfer typically involves multiple on-chain and off-chain operations distributed across several components, each of which can be attacked. We group these components into five functional domains (FDs), summarised in Table~\ref{tab:functional_domain_architecture} and Fig.~\ref{fig:Functional domain}. For the asset-transfer workflow abstracted in this study, FD-1 to FD-4 represent the principal processing stages; individual bridge designs may combine or reorder them. FD-5 is not a step in this path; it is a cross-cutting governance and operational control plane that manages how the other four domains operate.

\subsubsection*{FD-1: Deposit (Source Chain)}
The user starts a transfer on $\mathcal{S}$ by locking or burning an asset. The bridge contract checks the request, updates its records, and emits an event containing the transfer details, such as the sender, recipient, amount, and destination chain.

\subsubsection*{FD-2: Relay (Off-Chain)}
Off-chain agents, such as relayers, oracles, or validators, watch $\mathcal{S}$ for bridge events. Once a transaction is final, they collect its details, produce a proof or signed message, and transmit it to $\mathcal{D}$. FD-2 therefore produces and transmits the evidence, but does not decide whether it is acceptable.

\subsubsection*{FD-3: Verification (Destination Chain)}
A contract on $\mathcal{D}$ decides whether the evidence it receives is acceptable, that is, whether it shows a real action on $\mathcal{S}$. Common methods include checking block headers and Merkle proofs (light-client verification), requiring signatures from a minimum number of trusted validators (multi-signature verification), or checking a zero-knowledge proof~\cite{crosschain_19,crosschain_11}.

\subsubsection*{FD-4: Execution (Destination Chain)}
If verification succeeds, the bridge contract on $\mathcal{D}$ carries out the transfer by minting wrapped tokens, releasing locked assets, or calling another contract.

\subsubsection*{FD-5: Governance and Operations (Control Plane)}
This domain covers everyone and everything that manages the bridge over time, including privileged credentials, validator sets, multi-signature administrators, DAOs, timelocks, and upgrade mechanisms~\cite{crosschain_11}. It can change settings, upgrade contracts, pause the bridge, and replace validator keys. It also includes supporting infrastructure such as websites, DNS, cloud servers, and build systems. Because FD-5 governs or configures the other domains, a compromise of this layer can affect multiple stages of bridge operation.

\begin{table*}[pos=!tp]
\centering
\caption{FDs of a Cross-Chain Bridge and their Typical Weaknesses}
\label{tab:functional_domain_architecture}
\renewcommand{\arraystretch}{1.2}
\begin{adjustbox}{max width=\linewidth}
\begin{tabular}{p{2.3cm} p{2.8cm} p{3.4cm} p{3.0cm} p{4.4cm}}
\hline
\textbf{Domain} & \textbf{Role} & \textbf{Main Components} & 
\textbf{Security Objective} & \textbf{Typical Weaknesses} \\
\hline
\textbf{FD-1: Deposit} &
Lock or burn assets and record the transfer &
Source bridge contract, user wallet &
Integrity of the source-chain event &
Faulty locking logic, missing input checks, events emitted without a real deposit \\

\textbf{FD-2: Relay} &
Produce and transmit evidence of the transfer &
Relayers, oracles, off-chain listeners &
Integrity of the transmitted message &
Incorrect event interpretation, untrusted event sources, message-construction 
errors, relay compromise \\

\textbf{FD-3: Verification} &
Decide whether the evidence is acceptable &
Validators, multi-signature or MPC signers, proof-checking contracts &
Authenticity and validity of evidence &
Weak proof checks, wrong trust settings, forged signatures \\

\textbf{FD-4: Execution} &
Mint or release assets on the destination chain &
Destination bridge contract, liquidity pools &
Correct execution of approved transfers &
Incorrect state accounting, replayed claims, reentrancy, manipulable balances \\

\textbf{FD-5: Governance \& Operations} &
Manage and update the bridge over time &
Privileged credentials, validator sets, admin and upgrade rights, hosting and 
build systems &
Integrity of privileges and infrastructure &
Stolen or leaked keys, unauthorised upgrades, insecure DNS, cloud, or build systems \\
\hline
\end{tabular}
\end{adjustbox}
\end{table*}

\begin{figure*}[pos=!tp]
    \centering
    \includegraphics[width=\textwidth,max width=\linewidth]{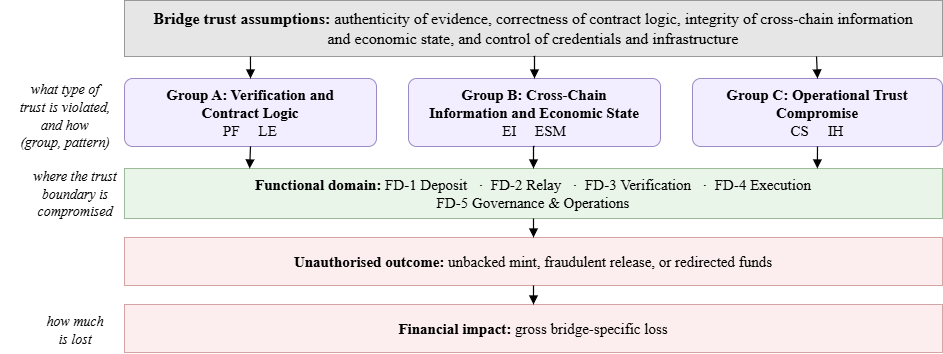}
    \caption{Proposed behaviour-based attack pattern architecture: each incident is classified by the type of trust assumption violated and how it is violated (group and attack pattern), where the compromised trust boundary lies (FD), and the resulting financial impact (gross bridge-specific loss).}
\label{fig:framework}
\end{figure*}

The attack patterns and the FDs are two separate dimensions: the pattern describes \emph{how} a trust assumption is violated, and the domain describes \emph{where} the compromised trust boundary lies. The domain is assigned to the primary location of the compromised trust boundary, not necessarily to the point where the final fraudulent action is executed. For 
incidents involving compromised credentials or infrastructure, FD-5 therefore captures the initial compromise of privileged control, even when FD-3 or FD-4 subsequently processes the resulting fraudulent action. Which domain a pattern lands in is decided incident by incident, not set beforehand. For example, incidents involving proof acceptance may map to FD-3, event-related manipulation may span FD-1 and FD-2, economic-state manipulation may affect FD-3 or FD-4 depending on the state the bridge uses, and logic flaws may occur in several domains. Fig.~\ref{fig:framework} summarises the resulting architecture, which Section~\ref{sec:patterns} presents in detail. Section~\ref{sec:analysis} reports the observed 
distribution across the dataset, and Section~\ref{sec:patterns} describes each pattern in detail, including real-world incidents and defences.

\section{Related Works}
\label{sec:related_work}

Cross-chain bridge security has been studied along four methodological 
directions: bridge architecture and trust models, attack taxonomies and security analysis, automated attack detection, and transaction-level and forensic analysis. Each direction addresses a different problem and classifies a different dimension of bridge security. This section summarises each direction, identifies the studies most relevant to ours, and explains how the present study is positioned relative to them.

\subsection{Cross-Chain Bridge Architecture and Trust Models}

Earlier work looked at blockchain interoperability and cross-chain mechanisms in general. Dong et al.~\cite{crosschain_15} survey blockchain technologies along with common attack vectors, and Li et al.~\cite{crosschain_17} cover mechanisms including notary schemes, hash-locks, sidechains, and relays. Cao et al.~\cite{crosschain_3} analyse the economic effects of bridges across 20 chains and 16 bridges. These studies establish key concepts of interoperability and trust models but do not systematically analyse real-world bridge attacks. Notland et al.~\cite{crosschain_11} provide a detailed architectural analysis of 64 bridges and 31 exploits across 13 components. This work motivates our functional-domain perspective; however, it focuses on architectural components and design flaws rather than incident-level functional classification.

\subsection{Bridge Attack Taxonomies and Security Analysis}
Several studies classify bridge attacks according to vulnerabilities, attack vectors, and exploit mechanisms. Belenkov et al.~\cite{crosschain_1} analyse major bridge hacks and their root causes, while Lee et al.~\cite{crosschain_20} review vulnerabilities in real-world bridge attacks. Li et al.~\cite{crosschain_2} classify bridge risks into smart-contract flaws, centralisation, liquidity, and oracle manipulation. Zhang et al.~\cite{crosschain_19} identify 12 attack vectors and 10 vulnerability types, while Zhang et al.~\cite{crosschain_18} define three vulnerability classes for automated detection. Mishra et al.~\cite{crosschain_12} and Werner et al.~\cite{crosschain_16} provide broader blockchain and DeFi security perspectives. These studies mainly classify attacks by vulnerability or mechanism and do not separately capture the violated trust assumption, functional domain, and financial impact at the incident level.

\subsection{Automated Detection and Monitoring}
Several tools now detect and monitor bridge attacks automatically. BridgeGuard~\cite{crosschain_4} collects 49 attacks and detects new ones from transaction execution graphs. XChainWatcher~\cite{crosschain_6} flags suspicious transactions with cross-chain logic rules, and BridgeShield~\cite{crosschain_9} and XScope~\cite{crosschain_18} monitor bridges using graph-based and rule-based methods respectively. Such tools show that automated detection works, but they set out to catch attacks, not to study systematically which trust they violate, which functional domain they hit, or how their financial impact develops over time.

\subsection{Empirical Transaction and Forensic Analysis}

Another line of work studies bridges at the level of individual transactions, looking at activity, behaviour, and how funds can be traced across chains. Yan et al.~\cite{crosschain_7} analyse large-scale cross-chain transactions and identify behavioural patterns, while XChainDataGen~\cite{crosschain_13} provides a large multi-chain transaction dataset. Pillai et al.~\cite{crosschain_10} trace Wormhole transaction flows, and forensic systems such as CONNECTOR~\cite{crosschain_8}, CONNEX~\cite{crosschain_5}, LOCARD~\cite{crosschain_14}, and the recent SoK by Zheng et al.~\cite{zheng2026sok} focus on linking, tracing, and matching cross-chain transactions using multi-source evidence. These studies demonstrate the importance of transaction-level and multi-source analysis, but mainly address transaction behaviour and forensic tracing rather than incident-level attack classification across a consistent longitudinal bridge corpus. Table~\ref{tab:related_work_categories} summarises these research directions.

\begin{table*}[pos=!tp]
\centering
\caption{Categorisation of Existing Cross-Chain Security Studies}
\label{tab:related_work_categories}
\renewcommand{\arraystretch}{1.2}
\begin{adjustbox}{max width=\linewidth}
\begin{tabular}{p{3.0cm} p{2.8cm} p{4.2cm} p{5.4cm}}
\hline
\textbf{Category} &
\textbf{Studies} &
\textbf{Research Focus} &
\textbf{Main Gap Relative to This Study} \\
\hline

Bridge Architecture \& Trust Models &
\cite{crosschain_3,crosschain_11,crosschain_17,crosschain_15} &
Architecture, components, interoperability, design flaws &
Architectural components and design flaws are analysed, but the studies do not 
use a mutually exclusive incident-level functional-domain coding scheme for 
longitudinal comparison \\

Attack Taxonomies \& Security Analysis &
\cite{crosschain_1,crosschain_2,crosschain_16,crosschain_19,crosschain_20} &
Vulnerability types, attack vectors, real exploits &
Primarily organised around vulnerability types, attack vectors, or 
architectural flaws rather than a separate incident-level 
violated-trust-assumption dimension, with limited longitudinal analysis \\

Automated Detection \& Monitoring &
\cite{crosschain_4,crosschain_6,crosschain_9,crosschain_12,crosschain_18} &
Attack detection and runtime monitoring &
Focused on detection accuracy rather than incident-level classification of 
patterns, domains, and losses \\

Transaction \& Forensic Analysis &
\cite{crosschain_5,crosschain_7,crosschain_8,crosschain_10,crosschain_13,crosschain_14,zheng2026sok} &
Transaction datasets, tracing, and measurement &
Transaction-level rather than incident-level, with limited integration of 
attack taxonomy and bridge functions \\

\hline
\end{tabular}
\end{adjustbox}
\end{table*}

\subsection{Research Gap and Positioning}

Existing cross-chain bridge studies examine architecture, trust models, attack types, and individual incidents~\cite{crosschain_4,crosschain_6,crosschain_11,crosschain_14,crosschain_19}. However, three gaps remain: (1) limited incident-level classification based on the \emph{trust assumption violated}; (2) insufficient separation between \emph{how} trust is violated and \emph{where} the trust boundary is compromised; and (3) lack of a consistently coded longitudinal corpus that combines trust-violation patterns, functional domains, and normalised financial losses. To address these gaps, this study analyses bridge attacks from January 2022 to August 2026 using a behaviour-based attack pattern architecture that combines a trust-violation-based classification, independent functional-domain mapping, explicit rules for ambiguous cases, and a common loss-normalisation rule.

\section{Data Collection and Methodology}
\label{sec:data_collection}

\begin{figure}[pos=!htbp]
    \centering
    \includegraphics[width=0.5\textwidth,max width=\linewidth]{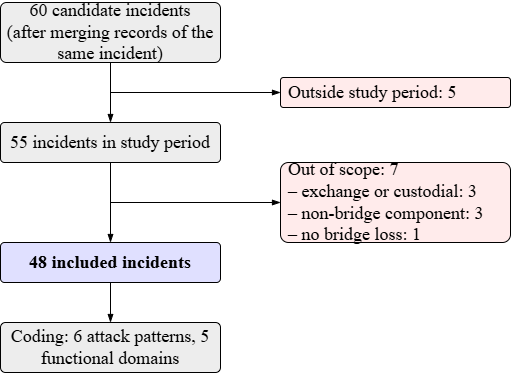}  
\caption{Dataset construction process.}
\label{fig:pipeline}
\end{figure}

This study systematically characterises real-world cross-chain bridge attacks by identifying recurring attacker behaviours and mapping those behaviours to the functional domains in which bridge trust assumptions are compromised. This section describes how the incident dataset was built and how each incident was classified. Fig.~\ref{fig:pipeline} summarises the process.

\subsection{Research Questions and Scope}

The study is guided by three research questions:

\begin{itemize}
    \item \textbf{RQ1:} Which attacker behaviours, understood as ways of breaking bridge trust assumptions, recur across real-world bridge attacks?
    \item \textbf{RQ2:} In which functional domains of a bridge do these behaviours strike?
    \item \textbf{RQ3:} Over the study period, how do attack patterns, functional domains, and financial losses change and spread over time?
\end{itemize}

The unit of analysis is a single attack on a cross-chain bridge or cross-chain 
messaging mechanism. The study period runs from 1 January 2022 to 31 August 
2026, based on the date each attack started. Repeat attacks on the same bridge at 
different times are recorded as separate incidents. An \emph{affected system} is 
the bridge or cross-chain protocol named in the incident record. Incident 
identity and system identity are distinct. Separate attacks occurring at 
different times are treated as separate incidents even when they affect the same 
system. Conversely, attacks affecting different components of the same protocol 
are assigned the same affected-system identifier; for example, Allbridge Core and 
the Allbridge CCTP router are counted as one system.

\subsection{Relation to Existing Incident Datasets}

Earlier studies have assembled collections of bridge attacks, including a 
taxonomy of 10 attack types and 12 attack vectors~\cite{crosschain_19} and 
49 incidents used to evaluate attack-transaction detection~\cite{crosschain_4}. 
The aim of this study is not a larger incident 
count. Its contribution lies in (i) an updated corpus covering January 2022 to 
August 2026 with a strict bridge-only inclusion rule, (ii) a behaviour-based attack 
pattern architecture grounded in the trust assumption the attacker breaks, (iii) a separate functional-domain 
mapping, (iv) explicit coding and boundary rules, (v) incident-level loss 
normalisation, (vi) a sensitivity analysis of ambiguous classifications, and 
(vii) a publicly auditable dataset.

\subsection{Incident Collection}

We collected candidate incidents from four types of sources, listed in 
Table~\ref{tab:sources}: peer-reviewed literature~\cite{crosschain_18, crosschain_19, crosschain_4, crosschain_6, crosschain_7, crosschain_13, crosschain_5, crosschain_14}, security company reports~\cite{halborn2022ronin, certik2022nomad, chainalysis2022wormhole, certik2023heco}, incident reports from the affected projects, and on-chain data. For the academic literature, we searched IEEE Xplore, ACM Digital Library, Scopus, Web of Science, and Google Scholar using combinations of the terms ``cross-chain bridge'', ``blockchain bridge'', ``cross-chain attack'', ``cross-chain security'', ``bridge exploit'', ``bridge hack'', ``bridge vulnerability'', ``bridge post-mortem'', ``message verification'', ``validator compromise'', ``relayer compromise'', and ``unauthorized asset transfer''. The same core terms were then run against security company reports, advisories, project post-mortems, and blockchain security write-ups; these sources helped us both find more incidents and check the reported details of each attack. To catch incidents that keyword searches missed, we went through the reference lists of related studies (backward citation searching) and public incident trackers. Where several records described one incident, we merged them before screening for eligibility, which left 60 candidate incidents.

\begin{table}[pos=!htbp]
\centering
\caption{Data Sources Used for Incident Collection}
\label{tab:sources}
\renewcommand{\arraystretch}{1.15}
\begin{adjustbox}{max width=\linewidth}
\begin{tabular}{p{2.2cm} p{2.6cm} p{2.6cm}}
\hline
\textbf{Source Type} & \textbf{Purpose} & \textbf{Examples} \\
\hline
Academic literature & Past incidents and existing classifications & 
SoKs and empirical studies \\
Security reports & Technical root cause & Halborn, CertiK, SlowMist, BlockSec \\
Project incident reports & Official timeline and cause & Post-mortems and fix commits \\
On-chain data & Transaction and loss checks & Block explorers, incident trackers \\
\hline
\end{tabular}
\end{adjustbox}
\end{table}

\subsection{Inclusion and Exclusion Criteria}

We \textbf{included} an incident only when the component that was exploited belonged to a cross-chain bridge or cross-chain messaging mechanism. That scope takes in deposit, lock, and burn handling; relaying; proof and message verification; release and mint execution; and the keys, governance, deployment, and operating infrastructure of the bridge itself.

We \textbf{excluded} any incident that was:
\begin{itemize}
    \item an attack on a centralised exchange or custodial platform;
    \item an attack on a non-bridge component of a protocol that also offers 
    bridging, such as its trading or lending contracts;
    \item an attack on an application that simply runs on several chains, or a 
    theft whose funds were only moved across chains afterwards;
    \item an incident with no loss to the bridge or its users; or
    \item an incident that started before 1 January 2022.
\end{itemize}

For inclusion purposes, loss requires a documented transfer, withdrawal, 
unauthorised minting, or other realised economic impact affecting 
bridge-controlled assets or bridge users. Attempts that were prevented before 
such an impact were excluded, so the corpus describes realised-loss bridge 
incidents rather than all attempted attacks.

Of the candidates, five lay outside the study period and seven fell out of scope (three exchange or custodial platforms, three non-bridge components, and one incident with no bridge loss), so 48 incidents remained. Each excluded incident, 
together with its sources and the reason for its exclusion, is listed in the 
public dataset.

\subsection{Evidence and Verification}

An incident was eligible when it had at least one of the following forms of 
support: (i) an official incident report or post-mortem from the affected 
project; (ii) at least two independent technical or reporting sources; or 
(iii) where no official root-cause disclosure existed, corroborated on-chain 
evidence together with independent third-party reporting. Each incident was then 
given an evidence level, defined in Table~\ref{tab:evidence}: route (i) 
corresponds to E1 or E3, depending on whether independent sources also exist; 
route (ii) to E2; and route (iii) to E4. Over 90\% of incidents are supported by 
an official report together with an independent source (E1) or by two or more 
independent sources (E2). The single E4 incident has no published root cause, so 
its classification is inferred from on-chain behaviour and is marked as such in 
the dataset. The levels describe the provenance of the evidence, not a strict 
ranking of its technical accuracy: a pair of detailed independent analyses (E2), for instance, can say more than a short official statement.

\begin{table}[pos=!htbp]
\centering
\caption{Evidence Levels of the 48 Incidents}
\label{tab:evidence}
\renewcommand{\arraystretch}{1.15}
\begin{adjustbox}{max width=\linewidth}
\begin{tabular}{c p{5.2cm} c}
\hline
\textbf{Level} & \textbf{Definition} & \textbf{Incidents} \\
\hline
E1 & Official post-mortem (or official fix) plus at least one independent 
technical source & 21 \\
E2 & At least two independent technical or reporting sources & 23 \\
E3 & Official post-mortem or incident report only & 3 \\
E4 & Third-party reports and on-chain tracing only; no official root-cause disclosure & 1 \\
\hline
\textbf{Total} & & \textbf{48} \\
\hline
\end{tabular}
\end{adjustbox}
\end{table}

\subsection{Loss Measurement}

Our loss measure is the \textit{gross bridge-specific loss}, meaning the value of assets taken from the bridge, its liquidity, or its users, or minted without backing, as a direct result of the incident, valued in US dollars at the time of the attack. We do not deduct later recoveries, voluntary returns, or bounty payments, so the figure reflects the gross financial impact of an incident and not what the attacker eventually kept. Left out are unrelated losses at the same organisation, damage the stolen assets later caused in other protocols, estimated market or price effects, and sums the attacker tried but failed to obtain. When sources gave a range instead of one value, we took the lower bound. We fixed this lower-bound rule in advance and applied it to every such case; it errs on the side of a smaller loss instead of guessing the most likely value, and it saves us from picking a point estimate the sources do not support. For attacks that lasted several days, the start date counts. All monetary values in the paper use this measure.

So that each value can be traced, the dataset stores three loss fields: the loss as first reported, the normalised gross bridge-specific loss that enters the analysis, and a note explaining any adjustment. A further \textit{loss basis} field says what the figure actually measures, namely bridge-held assets withdrawn, wallet funds drained, the value of unbacked minted assets, real assets acquired with unbacked tokens, or a reported amount with no stated basis.

\subsection{Dataset Structure}

The final dataset contains 48 evidence-supported incidents that meet the study's inclusion 
criteria. Each incident is recorded using the fields in Table~\ref{tab:schema}. 
The dataset, codebook and analysis scripts are available at 
\url{https://doi.org/10.5281/zenodo.23271026}.

\begin{table}[pos=!htbp]
\centering
\caption{Dataset Fields}
\label{tab:schema}
\renewcommand{\arraystretch}{1.15}
\begin{adjustbox}{max width=\linewidth}
\begin{tabular}{p{2.6cm} p{4.9cm}}
\hline
\textbf{Field} & \textbf{Description} \\
\hline
ID, Date, Year & Stable identifier (BR-001 to BR-048) and start date \\
Victim Platform & Affected system \\
Loss & Reported loss, gross bridge-specific loss (USD), normalisation note, 
and loss basis \\
Root Cause & Technical description of how the attack worked \\
Sources & Up to three evidence links \\
Evidence Level & E1 to E4 (Table~\ref{tab:evidence}) \\
Attack Pattern & One primary pattern (PF, LE, EI, ESM, CS, IH) \\
Functional Domain & One primary domain (FD-1 to FD-5) \\
Rationale & Reason for the pattern and domain labels \\
Boundary Case & Flag and alternative pattern and domain, if any \\
\hline
\end{tabular}
\end{adjustbox}
\end{table}

\subsection{Attack-Pattern Classification}

\textit{Taxonomy development and application.} Taxonomy development preceded 
the final coding of all incidents and followed four phases:

\begin{enumerate}
    \item \textbf{Concept generation:} Candidate categories were derived from
    recurring trust failures in prior cross-chain security literature and a
    preliminary review of the incidents.
    
    \item \textbf{Refinement:} Categories were refined to represent distinct
    violated trust assumptions with consistent boundary rules. For example,
    Flash Loan Amplification was replaced by Economic-State Manipulation.
    
    \item \textbf{Codebook fixing:} The six patterns, five domains, necessary
    conditions, and boundary rules were fixed in the codebook.
    
    \item \textbf{Coding:} The fixed codebook was applied to all 48 incidents.
\end{enumerate}

Section~\ref{sec:patterns} gives the full details.

\textit{Classification principle.}
Every bridge depends on trust assumptions, for example ``only authorised parties control signing keys'' or ``deposit events represent genuine deposits.'' The assumption that is broken defines the attack pattern, and the functional domain tells us which bridge subsystem was hit.

\textit{Primary pattern.}
To keep the counts free of double-counting, every incident gets exactly one primary pattern: the attacker behaviour that directly caused the compromise. Supporting methods such as flash loans, phishing, and social engineering are still recorded, but in the root-cause and rationale fields. We code a flash loan as ESM only if it was used to manipulate an economic state the bridge trusted.

\textit{Coding procedure.}
For each incident, we (1) extracted the attack mechanism, (2) identified the
violated trust assumption, (3) listed patterns whose necessary conditions were
met, (4) applied the boundary rules when multiple candidates remained,
(5) assigned the primary pattern and domain and recorded the rationale, and
(6) flagged boundary cases and their alternative labels. The same procedure
is summarized in Fig.~\ref{fig:decision_tree}.

\begin{table*}[pos=!tp]
\centering
\caption{Boundary Rules for Separating Attack Patterns}
\label{tab:boundary}
\renewcommand{\arraystretch}{1.15}
\begin{adjustbox}{max width=\linewidth}
\begin{tabular}{p{1.8cm} p{9.6cm} p{4.6cm}}
\hline
\textbf{Patterns} & \textbf{Boundary rule} & \textbf{Examples} \\
\hline
PF vs.\ LE &
A flawed or bypassable verifier $\rightarrow$ PF. A flaw in initialisation, an upgrade, or anything other than verification $\rightarrow$ LE. &
PF: ZKSwap, Verus, CrossCurve.
LE: Nomad, Ronin (2024), Messina. \\

PF vs.\ EI &
Forged or wrongly authenticated evidence $\rightarrow$ PF. Genuine evidence for a deposit/burn that is false $\rightarrow$ EI. &
PF: Pike Finance, Wanchain.
EI: Across, Coreum, warp.green. \\

CS vs.\ PF &
Stolen credentials approve the release or privileged changes themselves $\rightarrow$ CS. They just shift what the verifier trusts, letting forged evidence through $\rightarrow$ PF. &
CS: ALEX Lab, Seedify.
PF: GriffinAI, The Sandbox. \\

CS vs.\ IH &
Credentials taken from people, devices, storage, or key-management processes $\rightarrow$ CS. The bridge's build, deployment, or hosting systems are compromised $\rightarrow$ IH. &
CS: Ronin (2022), Harmony, THORChain.
IH: Force Bridge, Celer cBridge. \\

PF vs.\ IH &
The off-chain component that was compromised is the one that verifies $\rightarrow$ PF. &
PF: Kelp DAO. \\

ESM vs.\ LE &
A balance, price, or stake weight the bridge trusts is manipulated first $\rightarrow$ ESM. A coding flaw is exploited directly $\rightarrow$ LE. &
ESM: Allbridge (2023), Allbridge Core (2026), Shibarium.
LE: Messina, Li.Fi. \\

ESM vs.\ others &
When a flash loan merely pays for another exploit, the incident keeps that exploit's pattern. &
ESM: Allbridge (2023), Shibarium.
EI: Allbridge (2026, CCTP router). \\
\hline
\end{tabular}
\end{adjustbox}
\end{table*}

\subsection{Functional-Domain Mapping}

Each incident is mapped to one primary functional domain: the domain of the 
component whose trust was broken. Table~\ref{tab:domain_coding} gives the coding 
question used for each domain. The domain follows the component or trust boundary the attack compromised directly, not the later operation that the compromise made possible. An attack with stolen credentials therefore stays in FD-5 even when the credentials are afterwards used to approve verification or execution, since keys, validator sets, and operating infrastructure sit in the management layer.

\begin{table}[pos=!htbp]
\centering
\caption{Coding Questions for Functional-Domain Mapping}
\label{tab:domain_coding}
\renewcommand{\arraystretch}{1.15}
\begin{adjustbox}{max width=\linewidth}
\begin{tabular}{p{2.2cm} p{5.3cm}}
\hline
\textbf{Domain} & \textbf{Coding Question} \\
\hline
FD-1: Deposit & Did the attack tamper with the source-side record of a deposit, lock, or burn? \\
FD-2: Relay & Was the transmission or reading of a cross-chain event or message 
manipulated? \\
FD-3: Verification & Was a message or proof wrongly judged to be authentic or valid? \\
FD-4: Execution & Was an accepted instruction executed incorrectly, or based on 
manipulated state? \\
FD-5: Gov.\ \& Ops. & Were keys, validators, administrative rights, deployment, 
or infrastructure compromised? \\
\hline
\end{tabular}
\end{adjustbox}
\end{table}

\subsection{Reliability}

One author classified each incident and a second author checked it; where they disagreed, they talked it through. Since the two codings were not fully independent, we do not report an agreement score.

\subsection{Limitations of the Dataset}

The 48 incidents form a systematically built sample, not a full census of every bridge attack in the study period. Its main limitations are as follows:

\begin{itemize}
\item \textbf{Coverage and classification:} Smaller or thinly documented incidents may be missing from public reports, and for 2026 only January-August is covered. An attack can combine several behaviours, yet each incident carries a single primary pattern; one incident (TeleSwap, E4) had to be classified from scarce public evidence.

\item \textbf{Loss measurement:} The gross bridge-specific losses come from documented sources. Unbacked minting can be valued either by what was minted or by the real assets obtained; we record which basis was used and look at the alternatives where they apply.

\item \textbf{Generalisability:} With so few cases in some categories, conclusions about them are limited, and the taxonomy was built from the same corpus it is applied to. The results therefore speak to internal consistency, not external validity. Attacks that were prevented or caused no loss fall outside the scope.

\end{itemize}

We publish the codebook, the excluded incidents, the loss fields, and the evidence links so that others can check the work and build on it.

\section{Empirical Analysis}
\label{sec:analysis}

Here we turn to the 48 included incidents and the three research questions. Since every incident has a single primary attack pattern and a single functional domain (Section~\ref{sec:data_collection}), counts and losses add up without any double counting. All monetary figures are gross bridge-specific losses computed from unrounded incident-level values. The dataset is not a probability sample, so the results describe this corpus and should not be read as population-level attack rates. Because losses are heavily right-skewed, we give medians and interquartile ranges (IQRs) next to the aggregate figures.

\subsection{Temporal and Loss Distribution (RQ3)}
\label{sec:temporal}

Between January 2022 and August 2026 the dataset records 48 incidents across 43 distinct systems, with \$2.676 billion in total gross losses. A typical incident lost \$4.25M (median; IQR: \$1.18M-\$15.04M), far less than the mean of \$55.7M, which shows how concentrated the losses are.

January-August 2026 alone saw nineteen incidents, more than any full earlier year, each of which had between 3 and 10 (Table~\ref{tab:rate}). Still, 2026 is only partly observed and we have no exposure measures, such as how many bridges were active or how much liquidity they held, so this cannot be read as attacks becoming more frequent.
Median losses remained high in 2022-2023 (\$90.0M and \$86.8M) but fell
sharply to \$2.7-\$4.3M during 2024-2026. The year 2022 alone accounts for
70.8\% of total losses.

\begin{table*}[pos=!tp]
\centering
\caption{Incidents and Gross Losses per Year}
\label{tab:rate}
\renewcommand{\arraystretch}{1.15}
\footnotesize
\begin{adjustbox}{max width=\linewidth}
\begin{tabular}{c c c c c}
\hline
\textbf{Year} & \textbf{Months Observed} & \textbf{Incidents} &
\textbf{Total Loss} & \textbf{Median (IQR)} \\
\hline
2022 & 12 & 10 & \$1,894.8M & \$90.0M (3.3-292.0) \\
2023 & 12 & 3  & \$213.4M   & \$86.8M (43.7-106.4) \\
2024 & 12 & 8  & \$115.9M   & \$3.8M (1.5-11.7) \\
2025 & 12 & 8  & \$24.1M    & \$2.7M (1.2-4.1) \\
2026 & 8  & 19 & \$427.5M   & \$4.3M (0.7-9.9) \\
\hline
\end{tabular}
\end{adjustbox}
\par\smallskip
\parbox{\columnwidth}{\scriptsize Displayed values may not sum exactly to
totals because aggregates are calculated from unrounded incident-level losses.
2026 covers January-August only.}
\end{table*}

The temporal pattern also varies by attack type. The 2022 incidents were
distributed across five patterns, Logic Exploitation was most frequent in
2024, and Proof Forgery was most frequent in 2026, accounting for 8 of its
19 incidents. Credential Seizure is the only pattern observed in every year.

\begin{figure}[pos=!htbp]
    \centering
    \includegraphics[width=\columnwidth,max width=\linewidth]{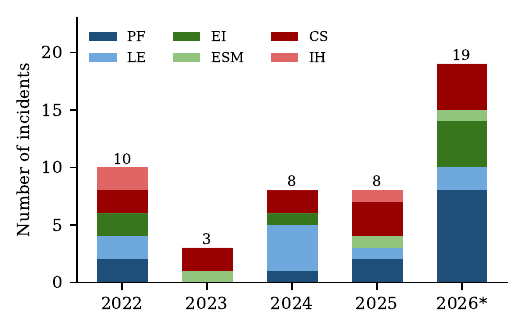}
    \caption{Number of incidents per year by attack pattern
    (*2026 covers January-August only).}
    \label{fig:incidents_year}
\end{figure}

\begin{figure}[pos=!htbp]
    \centering
    \includegraphics[width=\columnwidth,max width=\linewidth]{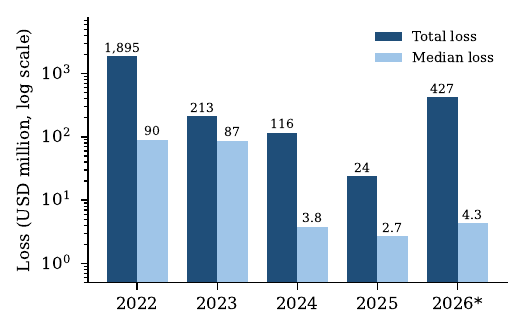}
    \caption{Total and median gross loss per year in USD million, with a
    logarithmic y-axis (*2026 covers January-August only).}
    \label{fig:loss_year}
\end{figure}

\subsection{Attack-Pattern Distribution (RQ1)}
\label{sec:pattern_distribution}

Proof Forgery (PF) and Credential Seizure (CS) are the most frequent patterns,
with 13 incidents each (Table~\ref{tab:pattern_dist}). Together, they account
for 54.2\% of incidents and 86.2\% of total losses. CS has the highest median
loss at \$10.7M, while Logic Exploitation (LE) ranks third in frequency but
accounts for only 10.1\% of total losses. The remaining three patterns
together account for 13 incidents and less than 4\% of losses.

\begin{table}[pos=!htbp]
\centering
\caption{Incidents and Gross Losses by Attack Pattern}
\label{tab:pattern_dist}
\renewcommand{\arraystretch}{1.15}
\footnotesize
\begin{adjustbox}{max width=\linewidth}
\begin{tabular}{l c c c}
\hline
\textbf{Pattern} & \textbf{Incidents (\%)} & \textbf{Total Loss (\%)} &
\textbf{Median (IQR)} \\
\hline
PF  & 13 (27.1) & \$1,233.4M (46.1) & \$7.00M (3.00-11.40) \\
LE  & 9 (18.8)  & \$270.4M (10.1)   & \$3.30M (1.60-12.00) \\
EI  & 7 (14.6)  & \$90.5M (3.4)     & \$1.30M (0.20-4.35) \\
ESM & 3 (6.3)   & \$4.6M (0.2)      & \$1.65M (1.11-2.02) \\
CS  & 13 (27.1) & \$1,072.3M (40.1) & \$10.70M (4.30-86.80) \\
IH  & 3 (6.3)   & \$4.3M (0.2)      & \$0.32M (0.28-2.04) \\
\hline
\textbf{Total} & 48 (100) & \$2,675.6M (100) &
\$4.25M (1.18-15.04) \\
\hline
\end{tabular}
\end{adjustbox}
\end{table}

\subsection{Functional-Domain Distribution (RQ2)}
\label{sec:domain_distribution}

Verification (FD-3) and Governance and Operations (FD-5) contain 31 of the
48 incidents (64.6\%) but account for 93.9\% of total losses
(Table~\ref{tab:domain_dist}). In contrast, Deposit and Relay (FD-1 and FD-2)
contain 8 incidents but only 3.4\% of losses. FD-5 is a cross-cutting control
plane rather than a processing stage and is therefore not directly comparable
with FD-1-FD-4.

\begin{table*}[pos=!tp]
\centering
\caption{Incidents and Gross Losses by Primary Functional Domain}
\label{tab:domain_dist}
\renewcommand{\arraystretch}{1}
\footnotesize
\begin{adjustbox}{max width=\linewidth}
\begin{tabular}{l c c c}
\hline
\textbf{Domain} & \textbf{Incidents (\%)} & \textbf{Total Loss (\%)} &
\textbf{Median (IQR)} \\
\hline
FD-1: Deposit       & 5 (10.4)  & \$84.7M (3.2)     & \$0.20M (0.19 - 4.20) \\
FD-2: Relay         & 3 (6.3)   & \$6.9M (0.3)      & \$1.30M (1.20 - 2.90) \\
FD-3: Verification  & 15 (31.3) & \$1,435.4M (53.6) & \$7.54M (3.00 - 101.00) \\
FD-4: Execution     & 9 (18.8)  & \$72.0M (2.7)     & \$2.40M (1.60 - 3.30) \\
FD-5: Gov.\ \& Ops. & 16 (33.3) & \$1,076.7M (40.2) & \$5.97M (1.08 - 82.83) \\
\hline
\textbf{Total} & 48 (100) & \$2,675.6M (100) &
\$4.25M (1.18 - 15.04) \\
\hline
\end{tabular}
\end{adjustbox}
\end{table*}

\subsection{Pattern - Domain Relationships (RQ2)}
\label{sec:pattern_domain}

\begin{figure}[pos=!htbp]
    \centering
    \includegraphics[width=\columnwidth,max width=\linewidth]{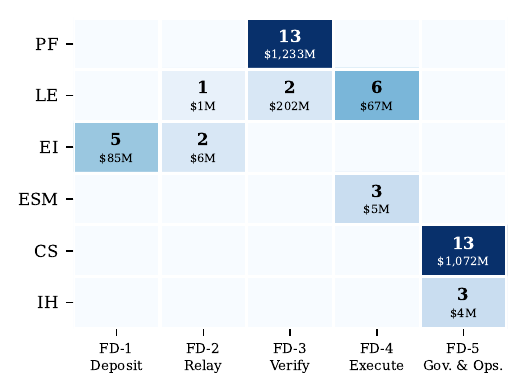}
    \caption{Number of incidents (bold) and total gross loss for each attack
    pattern and functional domain.}
    \label{fig:heatmap}
\end{figure}

Figure~\ref{fig:heatmap} cross-tabulates the two dimensions of the proposed
architecture: the attack pattern describes \emph{how} a trust assumption is
violated, while the functional domain identifies \emph{where} the compromised
trust boundary resides. PF is concentrated in FD-3, while CS and IH occur in
FD-5. All ESM incidents occur in FD-4, whereas EI spans FD-1 and FD-2 and LE
spans FD-2 - FD-4. The largest loss concentrations are PF - FD-3
(\$1,233M) and CS - FD-5 (\$1,072M). Because some mappings are constrained by
the taxonomy definitions, the heatmap is interpreted descriptively rather than
as evidence of an independent statistical association.

\subsection{Loss Concentration and Robustness}
\label{sec:loss_concentration}

Losses are highly concentrated, with a Gini coefficient of 0.83. The largest
incident accounts for 23.4\% of total losses, while the five and ten largest
incidents account for 74.7\% and 92.4\%, respectively. These results reinforce
the importance of reporting median losses alongside aggregate values.

To assess the robustness of the classification, all $2^6=64$ combinations of
the six boundary cases were evaluated. PF and CS remained the two most
frequent patterns in every combination, while their combined loss share ranged
from 75.3\% to 93.5\%. The FD-3 and FD-5 loss share remained approximately
94\%.

Loss-basis sensitivity produced a similar overall concentration but showed
that the relative contribution of PF depends partly on the treatment of
unbacked-mint valuations. Excluding the three major unbacked-mint valuations
reduces PF's loss share from 46.1\% to 3.3\%, while CS becomes the dominant
pattern with 71.9\% of losses. Nevertheless, the broader concentration of
losses in PF/CS and FD-3/FD-5 remains evident.

Four systems experienced multiple attacks, comprising nine incidents in total.
Verus-Ethereum and Allbridge experienced repeated attacks involving the same
pattern, whereas Ronin and Multichain were later attacked through different
patterns. Flash loans showed up in four incidents, where they mostly served to enable the attack instead of forming a pattern of their own.

\subsection{Summary of Empirical Findings}
\label{sec:empirical_summary}

All six patterns occur in the data, but the money lost is heavily concentrated. PF and CS cover 54.2\% of incidents and 86.2\% of losses, and FD-3 and FD-5 cover 64.6\% of incidents and 93.9\% of losses. Over time, 2026 shows the most incidents so far, while the typical loss per incident is much smaller than it was in 2022-2023. These results describe the corpus we observed; they say nothing about attack rates in the wider population.

\section{Proposed Behaviour-Based Attack Pattern Architecture}
\label{sec:patterns}
\label{sec:proposed_architecture_detailed}

This section presents the proposed architecture, defines the attack patterns,
describes their derivation and organisation, and specifies the classification
criteria used to code the dataset.

\subsection{Architecture Overview}

The proposed architecture (Fig.~\ref{fig:framework}) represents each incident
using four elements:

\begin{itemize}
    \item \textbf{Trust assumption:} the security condition on which the bridge
    relies, such as evidence authenticity or control of privileged keys.
    \item \textbf{Attack pattern:} \emph{how} the trust assumption is violated,
    represented by six patterns organised into three groups.
    \item \textbf{Functional domain:} \emph{where} the compromised trust
    boundary resides, using the five domains defined in Section~\ref{sec:preli}.
    \item \textbf{Financial impact:} the gross bridge-specific loss resulting
    from the incident.
\end{itemize}

Each incident receives one primary pattern and one primary functional domain,
allowing counts and losses to be aggregated without double counting. The
architecture was derived from the 48 incidents analysed in
Section~\ref{sec:analysis} and has not yet been validated on an independent
corpus.

\subsection{Taxonomy Definition}

\begin{definition}[Attack Pattern]
An attack pattern is a recurring trust-violation pattern defined by
(i) the bridge trust assumption that is broken, (ii) the mechanism used to
break it, and (iii) the resulting unauthorised cross-chain effect.
\end{definition}

The \emph{architecture} denotes the complete framework, while the
\emph{taxonomy} denotes its classification component. An \emph{attack
mechanism} is the concrete technical means used to execute an attack, whereas
a \emph{vulnerability} is the design or implementation weakness that enables
it. A \emph{functional domain} names the bridge component in which the compromised trust boundary sits.

We classify incidents, one at a time. Each one is given a single \emph{primary pattern}, the attacker behaviour that directly caused the compromise. Techniques that only supported the attack, such as phishing or flash loans, go into the root-cause and rationale fields and never add a second pattern label. Using one label avoids double counting, and the supporting mechanisms are not lost because the rationale still records them.

\subsection{Taxonomy Derivation}

We arrived at the patterns from the incident corpus in four steps:

\begin{itemize}
    \item \textbf{Incident coding:} before any label was given, we noted five attributes from the evidence, namely the attacker action, the enabling prerequisite, the bridge response, the broken trust assumption, and the unauthorised effect.
    \item \textbf{Candidate generation:} incidents sharing similar trust assumptions and attacker actions were put together as candidate patterns.
    \item \textbf{Boundary refinement:} we checked each candidate against the ambiguous incidents and kept it only if a clear rule could separate it from the others.
    \item \textbf{Final selection:} the criteria that survived were then used to code all 48 incidents.
\end{itemize}

\subsection{Taxonomy Hierarchy}

The six patterns are organised into three groups according to the type of trust
assumption violated (Fig.~\ref{fig:threemaincatagory}):

\begin{itemize}
    \item \textbf{Group A - Verification and Contract Logic:} whether the
    bridge correctly verifies and executes an action. PF breaks trust in
    verification, while LE breaks trust in other contract logic.
    \item \textbf{Group B - Cross-Chain Information and Economic State:} can the bridge rely on the information and economic state it works with? EI undermines deposit or burn signals, and ESM undermines balances, prices, or stake weights.
    \item \textbf{Group C - Operational Trust Compromise:} can the bridge rely on the people who run it and the systems behind it? CS undermines credential holders, and IH undermines infrastructure like hosting, DNS, or build systems.
\end{itemize}

\subsection{Classification Criteria}

Table~\ref{tab:classification_criteria} defines the necessary condition and
main exclusion rule for each pattern. The detailed boundary cases are given in
Table~\ref{tab:boundary}.

\begin{table*}[pos=!tp]
\centering
\caption{Classification Criteria for the Six Attack Patterns}
\label{tab:classification_criteria}
\renewcommand{\arraystretch}{1.15}
\begin{adjustbox}{max width=\linewidth}
\begin{tabular}{p{1.2cm} p{5.4cm} p{6.4cm} p{3.2cm}}
\hline
\textbf{Pattern} & \textbf{Necessary Condition} & \textbf{Assigned to Another
Pattern When} & \textbf{Dataset Examples} \\
\hline
PF & Evidence that is forged or invalid, and should fail the verification rule, is accepted by the verifier. &
The evidence is authentic but represents a nonexistent event (EI), or stolen
credentials directly approve the release (CS). &
Wormhole, BNB Chain Token Hub, Kelp DAO \\

LE & An unintended transfer is made possible directly by a coding flaw that lies outside verification. &
The attack requires prior manipulation of a trusted economic value (ESM). &
Nomad, Messina, Li.Fi \\

EI & A genuine signal is accepted although the deposit or burn it reports never happened or had no funds behind it. &
The evidence itself is forged or invalid and passes because verification is
flawed (PF). &
Qubit, Across, Coreum \\

ESM & An economic state value that the bridge trusts, and uses to approve or size a transfer, is manipulated by the attacker. &
A flash loan only funds another exploitable flaw. &
Allbridge (2023), Shibarium, Allbridge Core (2026) \\

CS & Genuine credentials are used by an unauthorised party to approve actions. &
The credentials are obtained through compromise of the bridge's own
infrastructure (IH). &
Ronin (2022), Multichain, AFX Trade \\

IH & The bridge's hosting, DNS, network, or build infrastructure is compromised. &
Credentials are directly compromised from people, devices, storage, or
key-management processes (CS), or the compromised component performs
verification (PF). &
EVODeFi, Celer cBridge, Force Bridge \\
\hline
\end{tabular}
\end{adjustbox}
\end{table*}

\textit{Key boundary rules.} The distinctions that matter most are listed below.

\begin{itemize}
    \item \textbf{PF vs.\ EI:} if forged or invalid evidence wrongly passes verification, the incident is PF; if the signal is genuine but misreports the economic event behind it, it is EI.
    \item \textbf{CS vs.\ PF:} stolen credentials that themselves approve the release or change control of the bridge make it CS; credentials that merely shift whom the verifier trusts, after which forged evidence gets through, make it PF.
    \item \textbf{CS vs.\ IH:} credentials compromised directly point to CS, while credentials taken by way of the bridge's own infrastructure point to IH.
    \item \textbf{PF vs.\ IH:} when the off-chain component that was compromised is the one doing the verification, the incident is PF.
    \item \textbf{ESM vs.\ LE:} ESM needs a trusted economic value to be manipulated first; if the coding flaw can be exploited straight away, the incident is LE.
\end{itemize}

\begin{figure}[pos=!htbp]
\centering
    \includegraphics[width=0.5\textwidth,max width=\linewidth]{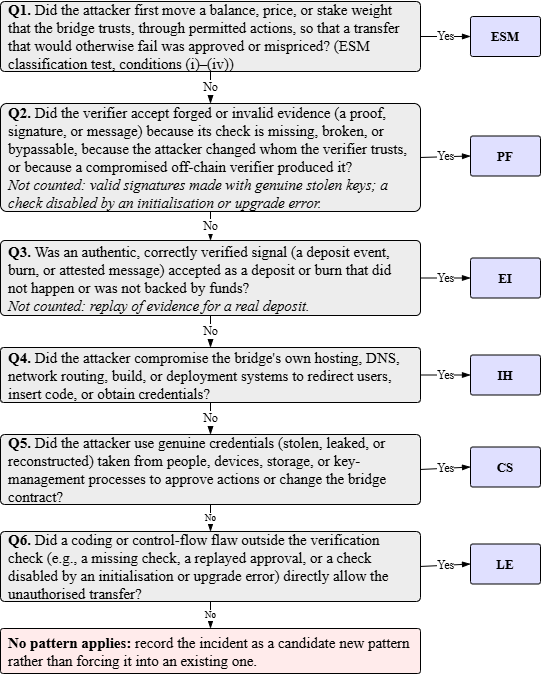}
\caption{Ordered decision procedure used to assign the primary attack pattern. The first condition that holds decides the pattern; the functional domain is assigned in a separate step.}
\label{fig:decision_tree}
\end{figure}

\textit{Ordered decision procedure.} In Figure~\ref{fig:decision_tree} the criteria are turned into a fixed sequence of questions, and the first question answered ``yes'' sets the primary pattern. The order only settles the boundary cases already defined; it adds no new criteria. The functional domain is assigned on its own, with Table~\ref{tab:domain_coding}.

\textit{Patterns versus domains.} We code pattern and domain separately. The pattern answers \emph{how} the trust assumption was broken, and the domain answers \emph{where} the compromised trust boundary is. By definition, some patterns sit in a single domain and others spread over several: PF, for example, maps to verification, while LE reaches several bridge components.

\subsection{Attack Patterns}

Each of the six patterns is a different way of breaking a bridge trust assumption. Below we give, for every pattern, its mechanism, the domains it affects, how common it is in the dataset, contributing factors, security implications, detection indicators, mitigations, and representative incidents. Losses are rounded to one decimal place from the incident-level data, and the public dataset holds the full evidence and rationale for every incident.

% GROUP A
\subsection*{Group A: Verification and Contract Logic}

Both patterns in this group break trust in how evidence is verified or how the bridge contracts behave. Proof Forgery goes after the verification mechanism itself; Logic Exploitation goes after contract logic that sits outside verification.

\subsubsection{Proof Forgery (PF)}

\begin{figure}[pos=!htbp]
    \centering
    \includegraphics[width=0.4\textwidth,max width=\linewidth]{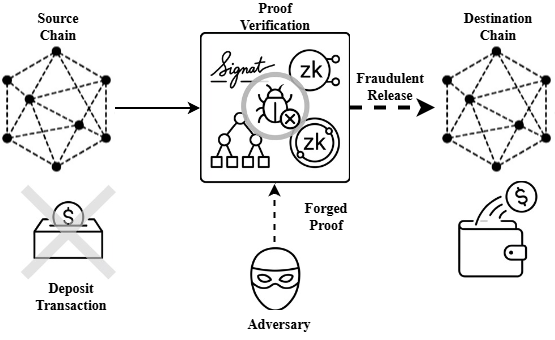}
    \caption{Proof Forgery attack pattern.}
    \label{fig:forgery}
\end{figure}

Here the attacker gets the verifier to accept forged or invalid evidence, whether a proof, a signature, or a message, and the result is an unbacked mint or release.

\textbf{Mechanism.}
What makes the attack possible is a verification flaw or a trust setting that can be changed. The attacker sends invalid evidence, the verifier lets it through, and the destination contract then releases or mints assets.

\textbf{Domain and prevalence.}
PF falls in FD-3 (Verification) and covers 13 incidents (27.1\%) and \$1,233.4M (46.1\%).

\textbf{Contributing factors.}
Complex verification logic, the fact that one invalid proof can be enough, and verifier or peer settings that can be changed all play a part.

\textbf{Implications and detection.}
None of the PF attacks we observed needed the attacker's own capital or privileged credentials, and a single forged message could free large sums. Warning signs are releases with no matching backed deposit and unexpected changes to verifier or peer settings.

\textbf{Mitigation.}
Verification logic should be audited and formally verified, backed by independent verifiers, real-time reconciliation between source and destination, rate limits, and emergency pauses.

\textbf{Representative incidents.}
Wormhole (BR-046, \$326M), BNB Chain Token Hub (BR-039, \$566M), and Kelp DAO
(BR-014, \$292M). Other PF incidents: BR-002, BR-005, BR-007, BR-011, BR-015,
BR-018, BR-019, BR-021, BR-025, and BR-032.

\subsubsection{Logic Exploitation (LE)}

\begin{figure}[pos=!htbp]
    \centering
    \includegraphics[width=0.4\textwidth,max width=\linewidth]{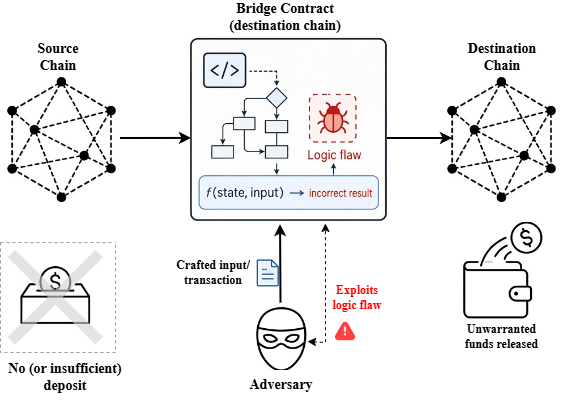}
    \caption{Logic Exploitation attack pattern.}
    \label{fig:logic}
\end{figure}

In this pattern, the attacker abuses contract logic outside verification, such as missing checks, replay protection, initialisation, or upgrade errors, with no forged evidence and no stolen keys.

\textbf{Mechanism.}
A bug in the code or in an upgrade opens a vulnerable execution path. A crafted input or sequence of transactions sets it off, and the contract carries out the unintended transfer.

\textbf{Domain and prevalence.}
LE spans FD-2 (Relay), FD-3 (Verification), and FD-4 (Execution), covering
9 incidents (18.8\%) and \$270.4M (10.1\%).

\textbf{Contributing factors.}
Contract code that anyone can inspect, contracts that are hard to patch, and security regressions introduced by upgrades all contribute.

\textbf{Implications and detection.}
The observed attacks did not require privileged credentials, and publicly
visible flaws could be rapidly reused. Signs to watch for are balance changes nobody can explain, claims made more than once, and security settings that disappear after an upgrade.

\textbf{Mitigation.}
Fuzzing and formal verification help, as do re-checking security settings after every upgrade, watching bridge balances, and pausing abnormal transfers automatically.

\textbf{Representative incidents.}
Nomad (BR-041, \$190M) and Messina (BR-016, \$47.5M). Other LE incidents:
BR-013, BR-026, BR-029, BR-030, BR-033, BR-034, and BR-048.

% GROUP B
\subsection*{Group B: Cross-Chain Information and Economic State}

Patterns in this group corrupt the information or economic state on which the bridge bases its cross-chain decisions.

\subsubsection{Event Injection (EI)}

\begin{figure}[pos=!htbp]
    \centering
    \includegraphics[width=0.48\textwidth,max width=\linewidth]{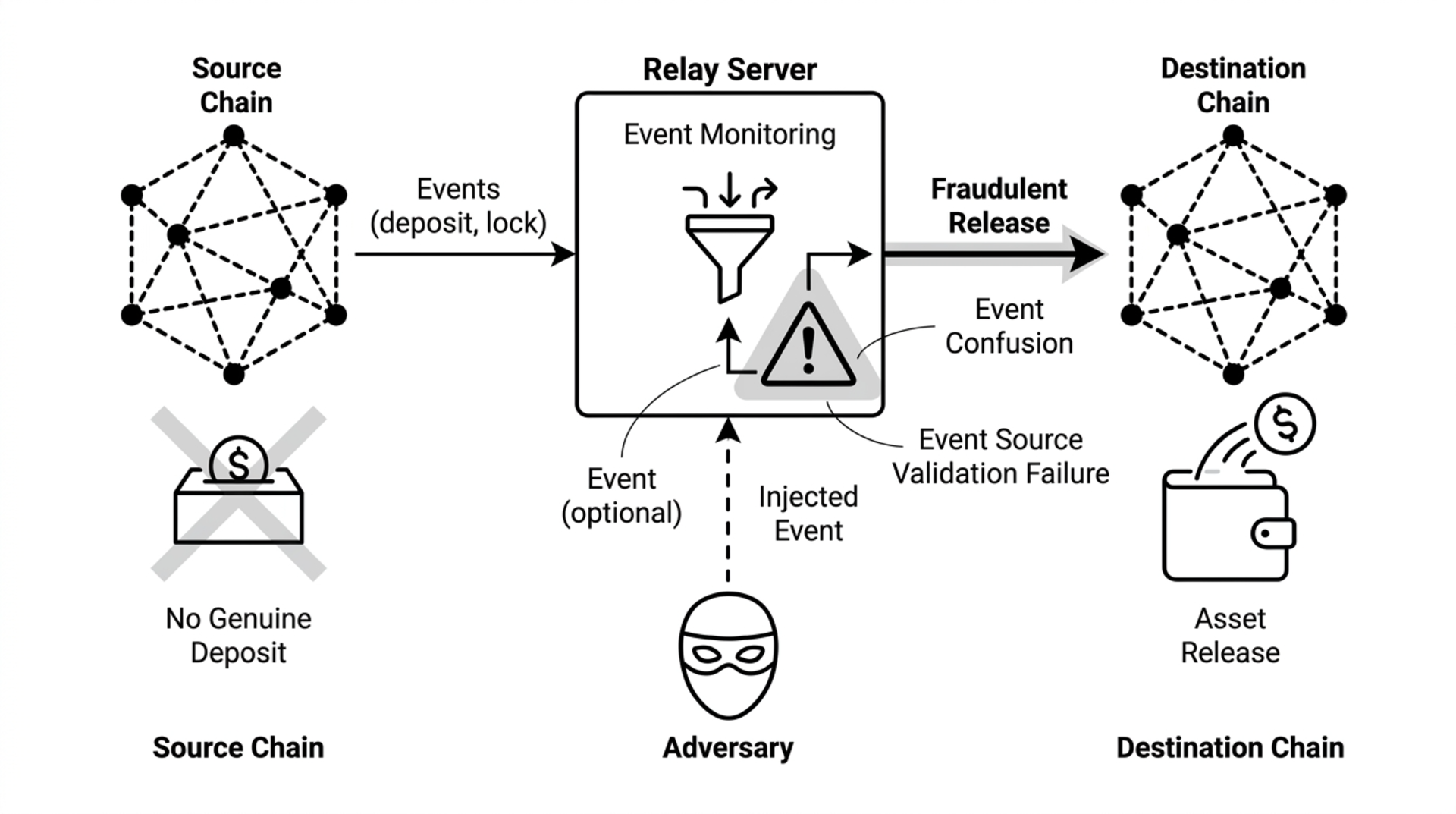}
    \caption{Event Injection attack pattern.}
    \label{fig:event_injection}
\end{figure}

The attacker produces, or gets accepted, a genuine signal for a deposit, burn, or other event that never happened or had no funds behind it. Verification itself works exactly as intended.

\textbf{Mechanism.}
Poor source validation or deposit handling lets a misleading signal be created. Relayers or validators pass it on, and the destination chain pays out unbacked assets.

\textbf{Domain and prevalence.}
EI occurs in FD-1 (Deposit) and FD-2 (Relay), with 7 incidents (14.6\%) and \$90.5M (3.4\%).

\textbf{Contributing factors.}
Weak checks on where a signal comes from, processing paths shared between token types, and relayers that fail together are the main contributors.

\textbf{Implications and detection.}
No privileged credentials were needed in the attacks we saw, and they could be run again and again until someone stopped them or the destination ran out of liquidity. The clearest sign is a deposit signal with no matching balance change on the source chain.

\textbf{Mitigation.}
Limit which sources may emit signals, confirm that funds have really arrived, run diverse relayer implementations, reconcile source balances, and stop relaying whenever they do not match.

\textbf{Representative incidents.}
Qubit (BR-047, \$80M) and Across (BR-009, \$4.5M). Other EI incidents:
BR-001, BR-003, BR-004, BR-028, and BR-045.

\subsubsection{Economic-State Manipulation (ESM)}

\begin{figure}[pos=!htbp]
    \centering
    \includegraphics[width=0.54\textwidth,max width=\linewidth]{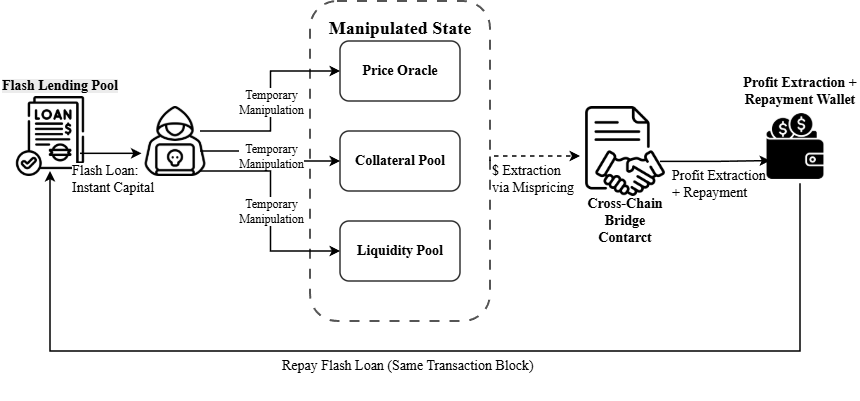}
    \caption{Economic-State Manipulation attack pattern.}
    \label{fig:flashloan}
\end{figure}

Here the attacker alters a value with economic meaning that the bridge trusts, such as a pool balance, an internal price, or a stake weight, and the bridge then relies on the altered value to approve or size a transfer that would otherwise be invalid. A flash loan only enables this; it is not the pattern itself.

\textbf{Classification test.}
We code an incident as ESM only if all of the following hold: (i) the bridge depends on an economic value that can be manipulated; (ii) the attacker shifts it using permitted actions; (iii) the shifted value feeds directly into the harmful transfer; and (iv) without the manipulation the transfer would fail. Condition (iv) is what distinguishes ESM from LE.

\textbf{Mechanism.}
Using swaps, deposits, or stake delegation, the attacker moves the trusted value for a short time, and the bridge then computes or approves an inflated transfer.

\textbf{Domain and prevalence.}
ESM affects FD-4 (Execution), covering 3 incidents (6.3\%) and \$4.6M (0.2\%).

\textbf{Contributing factors and implications.}
Because the bridge reads current values, and because liquidity can be borrowed briefly and executed atomically, manipulation can happen very quickly. Each of the three incidents we observed used a flash loan and was over within one transaction or a short execution window.

\textbf{Detection and mitigation.}
Look for sudden changes in balances or stake inside a single block together with briefly borrowed liquidity. Useful defences are time-averaged values, delays before new stake becomes active, monitoring of value changes, and circuit breakers.

\textbf{Scope.}
Only three incidents support ESM, so we treat it as provisional until future incidents confirm it.

\textbf{Representative incidents.}
Allbridge (BR-038, \$0.57M), Shibarium Bridge (BR-024, \$2.4M), and Allbridge
Core (BR-008, \$1.65M).

% GROUP C
\subsection*{Group C: Operational Trust Compromise}

In this group the attacker compromises the people, credentials, or infrastructure that the bridge trusts to keep it running.

\subsubsection{Credential Seizure (CS)}

\begin{figure}[pos=!htbp]
    \centering
    \includegraphics[width=0.4\textwidth,max width=\linewidth]{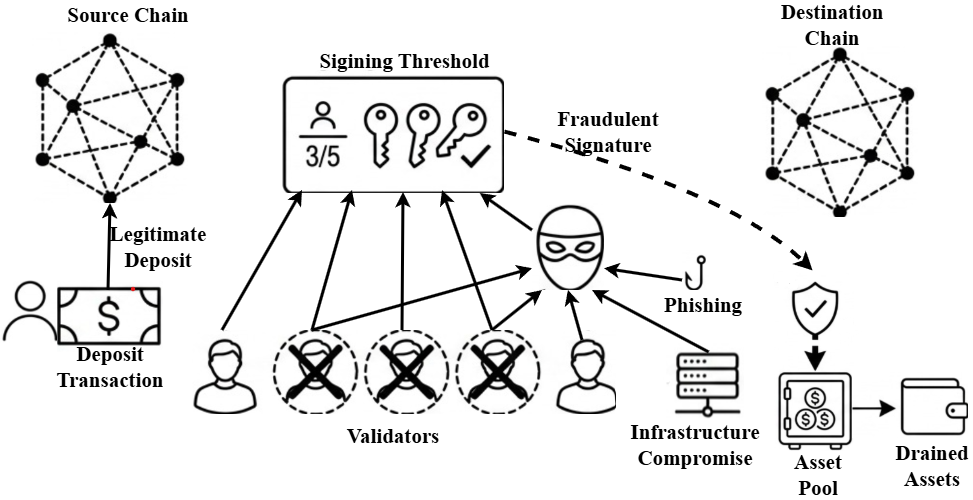}
    \caption{Credential Seizure attack pattern.}
    \label{fig:cred_seizure}
\end{figure}

The attacker gets hold of real validator, administrator, or deployer credentials straight from people, devices, storage, or key-management processes, and then uses them for actions that were never authorised.

\textbf{Mechanism.}
When keys are exposed or custody is concentrated, phishing, malware, leaks, or weak key-sharing can compromise them. The signatures produced are valid, so the bridge treats them as legitimate.

\textbf{Domain and prevalence.}
CS affects FD-5 (Governance \& Operations), covering 13 incidents (27.1\%) and
\$1,072.3M (40.1\%).

\textbf{Contributing factors.}
Keys held by too few parties, keys kept online, and the trade-off between security and day-to-day operability in threshold signing are the main factors.

\textbf{Implications and detection.}
The attacks we saw needed no contract flaw and could look like normal operations. CS had the highest median loss (\$10.7M). Unusual signing activity, upgrades, or permission changes are the signs to watch.

\textbf{Mitigation.}
Keep keys in HSMs or under MPC, spread custody over independent operators, watch signing activity, and require timelocks or challenge periods that are actively monitored.

\textbf{Representative incidents.}
Ronin Network (BR-043, \$625M), Multichain (BR-037, \$126M), and AFX Trade
(BR-006, \$24.2M). Other CS incidents: BR-010, BR-012, BR-017, BR-020,
BR-022, BR-023, BR-031, BR-035, BR-036, and BR-042.

\subsubsection{Infrastructure Hijacking (IH)}

\begin{figure}[pos=!htbp]
    \centering
    \includegraphics[width=0.48\textwidth,max width=\linewidth]{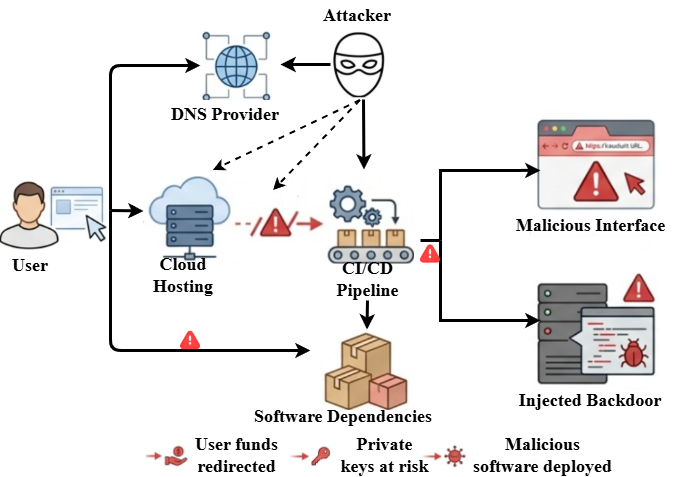}
    \caption{Infrastructure Hijacking attack pattern.}
    \label{fig:infra_hijack}
\end{figure}

Instead of attacking the bridge contracts, the attacker takes over hosting, DNS, network routing, or build systems in order to redirect users, slip in malicious code, or expose credentials.

\textbf{Mechanism.}
Once a weak operational system is compromised, users or operators keep using the altered system as if nothing had changed, and funds end up redirected or key material exposed.

\textbf{Domain and prevalence.}
IH affects FD-5 (Governance \& Operations), covering 3 incidents (6.3\%) and
\$4.3M (0.2\%).

\textbf{Contributing factors.}
Exposure grows with reliance on operational services, with dependencies hidden in the build pipeline, and with trade-offs between security and usability.

\textbf{Implications and detection.}
Contract audits often do not cover these systems, so a compromise can stay in place without touching any on-chain code. Unexpected changes to DNS, routing, certificates, or builds are the main warning signs.

\textbf{Mitigation.}
Apply least privilege, signed and reproducible builds, DNSSEC, continuous
domain and certificate monitoring, rapid interface shutdown, and key rotation.

\textbf{Scope.}
IH is supported by only three incidents and remains provisional, but it
represents a distinct risk outside the on-chain security boundary.

\textbf{Representative incidents.}
EVODeFi (BR-044, \$0.32M), Celer cBridge (BR-040, \$0.24M), and Force Bridge
(BR-027, \$3.76M).

% BOUNDARY CASES
\subsection{Boundary Cases}

Six incidents were used to test the most ambiguous classification boundaries.
Table~\ref{tab:boundary_cases} reports the assigned pattern and the decisive
criterion. The general classification rules are defined in
Table~\ref{tab:classification_criteria}.

\begin{table*}[pos=!tp]
\centering
\caption{Representative Boundary Cases}
\label{tab:boundary_cases}
\scriptsize
\renewcommand{\arraystretch}{1.1}
\begin{adjustbox}{max width=\linewidth}
\begin{tabular}{p{2.6cm} p{2.8cm} p{1.2cm} p{8.7cm}}
\hline
\textbf{Incident} & \textbf{Candidates} & \textbf{Assigned} & \textbf{Decisive criterion} \\
\hline
Nomad (BR-041) & PF, LE & LE &
An upgraded value let the logic check be skipped; neither forged evidence nor faulty proof verification was involved. \\

GriffinAI (BR-021) & CS, PF & PF &
The stolen key was used to swap the trusted peer, but the loss itself came from forged evidence that the verifier accepted. \\

Kelp DAO (BR-014) & IH, PF & PF &
The off-chain component that was compromised did the verification, so the trust that broke was in verification, not infrastructure. \\

Force Bridge (BR-027) & CS, IH & IH &
Malicious code planted in the deployment infrastructure was how the validator keys were obtained. \\

Shibarium Bridge (BR-024) & CS, ESM & ESM &
The withdrawal relied on briefly inflated stake; had the stake not been manipulated, the signing threshold would never have been met. \\

Allbridge CCTP router (BR-003) & EI, ESM, PF & EI &
The attestation was genuine; what went wrong was that the deposit was accepted before anyone checked that the funds had arrived. \\
\hline
\end{tabular}
\end{adjustbox}
\end{table*}

\section{Conclusion}
\label{sec:conclusion}
This paper presents an empirical analysis of 48 evidence-supported cross-chain bridge attacks from January 2022 to August 2026, resulting in \$2.676 billion in gross bridge-specific losses. We develop a behaviour-based attack pattern architecture grounded in bridge trust assumptions, comprising six attack patterns: \textit{Proof Forgery (PF)}, \textit{Logic Exploitation (LE)}, \textit{Event Injection (EI)}, \textit{Economic-State Manipulation (ESM)}, \textit{Credential Seizure (CS)}, and \textit{Infrastructure Hijacking (IH)} and map them across five functional domains: deposit, relay, verification, execution, and governance and operations. On the primary loss basis, Proof Forgery and Credential Seizure together explain 86.2\% of total losses, and the verification and governance and operations domains explain 93.9\%. A few very large incidents drive this concentration, but it remains when other loss bases are used. The results show that securing a bridge is not only about smart-contract bugs; trust assumptions, operational infrastructure, and governance matter just as much. Since the architecture comes from the same corpus it analyses, our next step is to apply the frozen codebook, with independent double coding, to incidents reported from now on, and to extend the framework to new cross-chain designs.

% \section*{Acknowledgements}
% No specific grant from any public, commercial, or not-for-profit funding agency supported this research.

\section*{CRediT authorship contribution statement}
\textbf{Ishraq Samiha}: Conceptualization, Methodology, Data curation, Formal analysis, Investigation, Visualization, Writing original draft; \textbf{Yasir Rahman}: Conceptualization, Methodology, Data curation, Formal analysis, Investigation, Visualization, Writing original draft; \textbf{Rahma Mukta}: Conceptualization, Methodology, Validation, Supervision, Writing  review \& editing.

\section*{Declaration of competing interest}
The authors have no conflicts of interest to declare.

\section*{Data availability}
The dataset, codebook and analysis scripts can be obtained from \url{https://doi.org/10.5281/zenodo.23271026}.

%bibliography style file
\bibliographystyle{cas-model2-names}

%bibliography database
\bibliography{references}

\end{document}